\documentstyle[epsf,12pt]{mrsproc}
\begin{document}
\begin{center}
{\bf Tunability of High-Dielectric-Constant Materials from First Principles}
\end{center}

\vspace{0.2cm}
\noindent
K. M. Rabe \\ 
Department of Physics and Astronomy, Rutgers University\\
Piscataway, New Jersey 08854-8019
\begin{abstract}
A first-principles method, based on density functional perturbation theory, is presented for computing the leading order tunability of
high-dielectric-constant materials.
\end{abstract}

\section{INTRODUCTION}
\label{sec:intro}

High-dielectric-constant materials are of great interest for high-performance
electronic devices. Besides the high dielectric constants themselves, which could in 
principle allow miniaturization of devices beyond that possible with silicon dioxide,
tunability of the dielectric response by a dc electrical bias opens up a 
wide range of additional applications.
Understanding the fundamental physics of high-dielectric-constant insulators
does, however, pose some significant challenges. In particular, the
effects of homogeneous electric fields on the electronic and crystal structure
of insulators, and on their dielectric and piezoelectric properties,
involves subtleties that have only recently begun to be systematically
addressed. Key advances include the correct formulation of the
electric field perturbation within density-functional perturbation theory (DFPT) \cite{baroni,gonze},
and the expression of polarization as a well-defined bulk property through
the Berry-phase formalism \cite{ksv}. Subsequently, the effects of finite fields have
been formulated using Taylor expansions around zero field \cite{nunes}. Using 
a closely related approach, a formalism in which polarization is the independent
variable has been developed \cite{sai} which includes a method for computing
the nonlinear dielectric susceptibility.

In this paper, we present a first-principles method for the efficient direct computation of 
the tunability of high-dielectric-constant materials to leading order in the
electric field. This method is designed for implementation into existing
density functional pertubation theory packages, such as ABINIT~\cite{ABINIT} and PWSCF/PHONON~\cite{PWSCF},
which typically calculate second derivatives of the energy and can be extended
to compute third derivatives without significant additional computational 
effort. If third derivatives are available, the collinear susceptibility of a ferroelectric material
can be computed with the sole approximation that the electric-field derivative
of the electronic susceptibility be fixed at the value at zero field with corresponding
equilibrium structural parameters.
For noncollinear tunability or
tunability of paraelectrics, or if the DFPT packages are to be used for 
collinear susceptibility of ferroelectrics without extension to computation
of third derivatives, nearly all of the necessary higher derivatives can be
obtained from finite differences with a calculation at {\it a single} additional 
structure. However, to complete the calculation in these cases, it is necessary 
to estimate the highest derivatives of the electronic susceptibility
with respect to the structural parameters, as their complete specification would
require additional finite-difference calculations.
These approximations should be reasonable for the materials of interest, 
in which the dielectric constant is dominated by the large lattice
contribution.

\section{FORMALISM}
\label{sec:forma}

We consider the dielectric response of a polar insulator to a homogeneous static electric field $\cal E$.
The first steps of the analysis are identical to those presented in Ref. \cite{sai}.
Given a particular unit cell, let ${\bf R}$ be the atomic coordinates and $\eta$ the homogeneous strain relative to a suitable reference structure. 
The total energy $E({\bf {\cal E}})$ and equilibrium structure ${\bf R}(\cal E)$ and $\eta(\cal E)$ of the system in the field $\cal E$ are obtained by the minimization
\begin{equation}
\min_{{\bf R},\eta} {E({\bf R},\eta,{\bf {\cal E}})} 
\label{eq:minE}
\end{equation}
with ${\bf R}(\cal E)$, $\eta(\cal E)$, and
$E({\bf {\cal E}})$ being the values at the minimum.
The polarization ${\bf P}({\bf R},\eta,\cal E )$, the thermodynamic conjugate of $\cal E$, can then be obtained from the expression 
${\bf P}({\bf R},\eta,{\cal E}) = -{\partial E({\bf R},\eta, {\cal E})\over \partial {\cal E}}\vert_{{\bf R},\eta}$ and
${\bf P}(\cal E )$ obtained by evaluating ${\bf P}({\bf R}({\cal E}),\eta({\cal E}),{\cal E})$, or equivalently,
-${d E({\cal E})\over d{\cal E}}$.

In DFT, there is as yet no rigorous formulation for finite $\cal E$, so we expand in $\cal E$ around $\cal E$ = 0:
\begin{eqnarray}
E({\bf R},\eta,{\bf {\cal E}}) 
&=& E({\bf R},\eta, {\bf {\cal E}}=0)+
{\bf {\cal E}}_\alpha E^{(1)}_{\alpha}({\bf R},\eta) \nonumber\\
&+&\frac{1}{2}{\cal E}_\alpha {\cal E}_\beta 
E^{(2)}_{\alpha\beta}({\bf R},\eta)
+\frac{1}{6}{\bf {\cal E}}_\alpha {\bf {\cal E}}_\beta {\bf {\cal E}}_\gamma
E^{(3)}_{\alpha\beta\gamma}({\bf R},\eta)+
\cdots 
\label{eq:Eexp}
\end{eqnarray}
where we have introduced the compact notation $$E^{(n)}_{\alpha_1\cdots\alpha_n}({\bf R},\eta)=
\left.\frac{\partial^n E({\bf R},\eta,{\cal E})}{\partial {\cal E}_{\alpha_1}\cdots\partial{\cal E}_{\alpha_n}}\right|_{{\cal
E}=0}.$$
Carried to all orders in $\cal E$, this expansion is exact, but for practical purposes at sufficiently small fields we truncate this sum to define $E_i({\bf R},\eta,{\bf {\bf {\cal E}}})$ as the sum of the
first $i+1$ terms in Eq.~(\ref{eq:Eexp}). We will also use similar expansions of ${\bf R}({\cal E})$ and
$\eta({\cal E})$. ${\bf R}_0$ and $\eta_0$ are the zero-field equilibrium structural parameters, routinely computed by
using first-principles forces and stresses, and are assumed, along with the spontaneous polarization ${\bf P}_s={\bf P}({{\bf R}_0,\eta_0,0})$, to be known at the beginning of the tunability calculation.

We first consider the collinear tunability of a polar insulator with a nonzero spontaneous polarization ${\bf P}_s$ 
in a field $\cal E$ parallel to ${\bf P}_s$, taken to define the $\hat z$ direction.
This allows us to treat $\cal E$, $P$, $\chi_\infty$ and $\chi'_\infty$ as scalars,
where $\chi_\infty({\bf R},\eta)=E^{(2)}_{zz}({\bf R},\eta)$ and
$\chi'_\infty({\bf R},\eta)=E^{(3)}_{zzz}({\bf R},\eta)$.  
To clarify the logic of the presentation, we will write the expressions that follow for a single scalar $R$ and a single component of the strain
tensor $\eta$; the generalization to the full set of atomic coordinates and strain tensor components 
is straightforward.

We first assume that for any given structure, the DFPT package has the capability of calculating
all third derivatives of the energy with respect to $R$, $\eta$ and $\cal E$.
This leads us to consider E$_3$ and approximate $\chi'_\infty({R},\eta)\approx \chi'_\infty({R_0},\eta_0)$,
resulting in the following key expressions:
\begin{eqnarray}
E_3({R},\eta,{\cal E}) &=& E({R},\eta,0)-P({R},\eta,0){\cal E}-\frac{1}{2}\chi_\infty({R},\eta){\cal E}^2-\frac{1}{6}\chi'_\infty({R}_0,\eta_0){\cal E}^3\\
P({R},\eta,{\cal E}) &=& P({R},\eta,0)+{\cal E}\chi_\infty({R},\eta)+\frac{1}{2}\chi'_\infty({R}_0,\eta_0){\cal E}^2.
\end{eqnarray}

Our interest is in the $\cal E$ derivative of the static dielectric susceptibility 
$${d\chi_s({\cal E})\over d{\cal E}}={d^2 P({\cal E})\over d{\cal E}^2} ={d^2P({{R}({\cal E}),\eta(\cal E),\cal E}) \over d{\cal E}^2}$$
correct to leading order in the field. In the case under consideration, the zeroth order term suffices.
This is a total derivative with respect to $\cal E$, and thus involves derivatives of P with respect to R and $\eta$
as well as $\cal E$, as well as derivatives of ${R}({\cal E})$ and $\eta({\cal E})$ at zero field, as follows:
\begin{eqnarray}
{d\chi_s({\cal E})\over d{\cal E}}\vert_0&=&
\chi'_\infty(R_0,\eta_0)+
2\left.\frac{\partial \chi_\infty}{\partial R}\right|_{0}R^{(1)}+
2\left.\frac{\partial \chi_\infty}{\partial \eta}\right|_{0}\eta^{(1)}\nonumber\\
&+&\left.\frac{\partial^2 P}{\partial R^2}\right|_{0}{R^{(1)}}^2+
\left.\frac{\partial^2 P}{\partial \eta^2}\right|_{0}{\eta^{(1)}}^2+
2\left.\frac{\partial^2 P}{\partial R \partial \eta}\right|_{0}R^{(1)}\eta^{(1)}+
\left.\frac{\partial P}{\partial R}\right|_{0}R^{(2)}+
\left.\frac{\partial P}{\partial \eta}\right|_{0}\eta^{(2)}
\label{eq:tunability}
\end{eqnarray}

For given $\cal E$, the equilibrium structural parameters ${R}({\cal E})$ and $\eta({\cal E})$ 
are found as solutions of the extremum equations:
\begin{eqnarray}
0&=&\frac{dE_3({R},\eta,{\cal E})}{dR}=
\frac{\partial E({R},\eta,0)}{\partial R}
-{\cal E}\frac{\partial P({R},\eta,0)}{\partial R}
-\frac{{\cal E}^2}{2}\frac{\partial \chi_\infty({R},\eta)}{\partial R}
+\cdots\\
0&=&\frac{dE_3({R},\eta,{\cal E})}{d\eta}=
\frac{\partial E({R},\eta,0)}{\partial \eta}
-{\cal E}\frac{\partial P({R},\eta,0)}{\partial \eta}
-\frac{{\cal E}^2}{2}\frac{\partial \chi_\infty({R},\eta)}{\partial \eta}
+\cdots
\end{eqnarray}
We expand ${R}({\cal E})=R_0 + R^{(1)}{\cal E} + \frac{1}{2}R^{(2)}{\cal E}^2$ and $\eta({\cal E})=\eta_0 + \eta^{(1)}{\cal E} + \frac{1}{2}\eta^{(2)}{\cal E}^2$.
By solving the extremum equations to linear order in $\cal E$, $R^{(1)}$ and $\eta^{(1)}$ can be readily obtained in terms of second derivatives evaluated at the zero-field equilibrium structure:
\begin{eqnarray}
\pmatrix{R^{(1)}\cr \eta^{(1)}\cr}=
\pmatrix{\frac{\partial^2 E}{\partial R^2} & 
\frac{\partial^2 E}{\partial R \partial \eta}\cr 
\frac{\partial^2 E}{\partial \eta \partial R }& 
\frac{\partial^2 E}{\partial \eta^2}\cr}^{-1}
\pmatrix{\frac{\partial P}{\partial R}\cr 
\frac{\partial P}{\partial \eta}\cr}
\label{eq:R1}
\end{eqnarray}
At quadratic order, we similarly obtain $R^{(2)}$ and $\eta^{(2)}$. 
\begin{eqnarray}
\pmatrix{R^{(2)}\cr \eta^{(2)}\cr} =
\pmatrix{\frac{\partial^2 E}{\partial R^2} & 
\frac{\partial^2 E}{\partial R \partial \eta}\cr 
\frac{\partial^2 E}{\partial \eta \partial R }& 
\frac{\partial^2 E}{\partial \eta^2}\cr}^{-1} 
\pmatrix{D_R \cr D_\eta \cr}
\label{eq:R2}
\end{eqnarray}
where
\begin{eqnarray}
D_R=2\frac{\partial^2 P}{\partial R^2}R^{(1)}+
2\frac{\partial^2 P}{\partial R \partial \eta}\eta^{(1)}+
\frac{\partial \chi_\infty}{\partial R}
-\frac{\partial^3 E}{\partial R^3}{R^{(1)}}^2
-2\frac{\partial^3 E}{\partial \eta \partial R^2}R^{(1)}\eta^{(1)}
-\frac{\partial^3 E}{\partial \eta^2 \partial R}{\eta^{(1)}}^2\\
D_\eta=2\frac{\partial R \partial \eta}{\partial R^2}R^{(1)}+
2\frac{\partial^2 P}{\partial \eta^2}\eta^{(1)}+
\frac{\partial \chi_\infty}{\partial \eta}
-\frac{\partial^3 E}{\partial \eta \partial R^2}{R^{(1)}}^2
-2\frac{\partial^3 E}{\partial \eta^2 \partial R}R^{(1)}\eta^{(1)}
-\frac{\partial^3 E}{\partial \eta^3}{\eta^{(1)}}^2
\end{eqnarray}
Inserting all these results into Eq.~\ref{eq:tunability} completes the calculation of the linear dependence of $\chi_s$ on electric field.

Next, we assume that 
the highest derivatives of the energy that are available from
the DFPT package used are the second derivatives \cite{nostrain}. 
For example, we can directly compute $\frac{\partial P}{\partial R}\vert_0$ but not $\frac{\partial^2 P}{\partial R^2}\vert_0$. As discussed in more detail below, a finite difference trick allows us to obtain relevant combinations of third derivatives one order higher in $R$ or $\eta$, but not in $\cal E$ so that under these conditions, $\chi'_\infty(R,\eta)$, the electric field dependence of the electronic susceptibility, is not available. 
However, it is reasonable to neglect it, as the electronic susceptibility is small relative to the lattice contribution.
To determine $R^{(2)}$ and $\eta^{(2)}$, we also need $\frac{\partial \chi_\infty}{\partial R}\vert_0$ and $\frac{\partial \chi_\infty}{\partial \eta}\vert_0$. From finite difference, we will be able
to determine only the combination $\frac{\partial \chi_\infty}{\partial R}\vert_0 R^{(1)}+
\frac{\partial \chi_\infty}{\partial \eta}\vert_0 \eta^{(1)}$. We obtain an estimate by assuming that all terms contribute equally to the sum.

We first do a DFPT calculation at $R_0$, $\eta_0$ including all second derivatives. From this, we can 
obtain $R^{(1)}$ and $\eta^{(1)}$ from Eq.~\ref{eq:R1}.
Next, we chose a small field ${\cal E}_1$ and do a second DFPT calculation at structure $R_1=R_0+R^{(1)}{\cal E}_1$ and
$\eta_1=\eta_0+\eta^{(1)}{\cal E}_1$. Using finite difference expressions, we can obtain the necessary combinations
of third derivatives at the zero field equilibrium structure, as follows.
For example, we need the $\cal E$ dependence of $\chi_\infty({R}({\cal E}),\eta({\cal E}))$ only to linear order. 
The coefficient of the linear term is 
\begin{equation}
\frac{\partial \chi_\infty}{\partial R}\vert_0 R^{(1)}+
\frac{\partial \chi_\infty}{\partial \eta}\vert_0 \eta^{(1)}
\label{eq:linearchi}
\end{equation}
Since $R_1({\cal E})-R_0=R^{(1)}{\cal E}$ and $\eta_1({\cal E})-\eta_0=\eta^{(1)}{\cal E}$ are linear in $\cal E$, we expand $\chi_\infty({R},\eta)$ around $R_0$, $\eta_0$ to linear order in $R-R_0$ and $\eta-\eta_0$ and evaluate at $R_1$, $\eta_1$.
$$\chi_\infty({R},\eta)=\chi_\infty({R}_0,\eta_0)+
\frac{\partial \chi_\infty}{\partial R}\vert_0 R^{(1)}(R-R_0)+
\frac{\partial \chi_\infty}{\partial \eta}\vert_0 \eta^{(1)}(\eta-\eta_0)$$
to express the combination of derivatives given by Eq.~\ref{eq:linearchi} as
$$(\chi_\infty({R}_1,\eta_1)-\chi_\infty({R}_0,\eta_0))/{{\cal E}_1})$$


The calculation of P is a little more involved, since we need the $\cal E$ dependence to quadratic order.
\begin{eqnarray}
&&P({R}({\cal E}),\eta({\cal E}),0)=P({R}_0,\eta_0,0)+
{\cal E}(\frac{\partial P}{\partial R}\vert_0 R^{(1)} + \frac{\partial P}{\partial \eta}\vert_0 \eta^{(1)})\nonumber\\
&+&\frac{1}{2}{\cal E}^2(\frac{\partial P}{\partial R}\vert_0 R^{(2)} + \frac{\partial P}{\partial \eta}\vert_0 \eta^{(2)}
+\frac{\partial^2 P}{\partial R^2}\vert_0 {R^{(1)}}^2 + \frac{\partial^2 P}{\partial \eta^2}\vert_0 (\eta^{(1)})^2 
+2\frac{\partial^2 P}{\partial R \partial \eta}\vert_0 R^{(1)}\eta^{(1)})
\end{eqnarray}
The linear coefficient is a combination of available derivatives with $R^{(1)}$ and $\eta^{(1)}$.
The two contributions to the quadratic coefficient (first derivatives of P and second derivatives of P)
are determined by separate calculations.
First, we use
\begin{eqnarray}
P(R_1,\eta_1,0)&=&P({R}_0,\eta_0,0)
+{\cal E}_1(\frac{\partial P}{\partial R}\vert_0 R^{(1)} + \frac{\partial P}{\partial \eta}\vert_0 \eta^{(1)})\nonumber\\
&+&\frac{1}{2}{\cal E}_1^2(\frac{\partial^2 P}{\partial R^2}\vert_0 {R^{(1)}}^2 + \frac{\partial^2 P}{\partial \eta^2}\vert_0 (\eta')^2 
+2\frac{\partial^2 P}{\partial R \partial \eta}\vert_0 R^{(1)}\eta')
\end{eqnarray}
to obtain
\begin{eqnarray}
&&\frac{\partial^2 P}{\partial R^2}\vert_0 {R^{(1)}}^2 + \frac{\partial^2 P}{\partial \eta^2}\vert_0 (\eta^{(1)})^2 
+2\frac{\partial^2 P}{\partial R \partial \eta}\vert_0 R^{(1)}\eta^{(1)}\nonumber\\
&=&\frac{2}{{\cal E}_1^2}(P(R_1,\eta_1,0)-P({R}_0,\eta_0,0)
-{\cal E}_1(\frac{\partial P}{\partial R}\vert_0 R^{(1)} + \frac{\partial P}{\partial \eta}\vert_0 \eta^{(1)}))
\end{eqnarray}
Then finally, we have to express $R^{(2)}$ and $\eta^{(2)}$ (from Eq.~\ref{eq:R2}) in terms of available derivatives.
This involves using finite difference expressions 
\begin{eqnarray}
\left.\frac{\partial P({R},\eta,0)}{\partial R}\right|_{1}=
\left.\frac{\partial P({R},\eta,0)}{\partial R}\right|_{0}
+{\cal E}_1(\left.\frac{\partial^2P({R},\eta,0)}{\partial R^2}\right|_{0}R^{(1)}
+\left.\frac{\partial^2P({R},\eta,0)}{\partial R \partial \eta}\right|_{0}\eta^{(1)})\\
\left.\frac{\partial P({R},\eta,0)}{\partial \eta}\right|_{1}=
\left.\frac{\partial P({R},\eta,0)}{\partial \eta}\right|_{0}
+{\cal E}_1(\left.\frac{\partial^2P({R},\eta,0)}{\partial R \partial \eta}\right|_{0}R^{(1)}
+\left.\frac{\partial^2P({R},\eta,0)}{\partial \eta^2}\right|_{0}\eta^{(1)})\\
\left.\frac{\partial^2 E({R},\eta,0)}{\partial R^2}\right|_{1}=
\left.\frac{\partial^2 E({R},\eta,0)}{\partial R^2}\right|_{0}
+{\cal E}_1(\left.\frac{\partial^3E({R},\eta,0)}{\partial R^3}\right|_{0}R^{(1)}
+\left.\frac{\partial^3E({R},\eta,0)}{\partial R^2 \partial \eta}\right|_{0}\eta^{(1)})\\
\left.\frac{\partial^2 E({R},\eta,0)}{\partial R \partial \eta}\right|_{1}=
\left.\frac{\partial^2 E({R},\eta,0)}{\partial R \partial \eta}\right|_{0}
+{\cal E}_1(\left.\frac{\partial^3E({R},\eta,0)}{\partial R^2 \partial \eta}\right|_{0}R^{(1)}
+\left.\frac{\partial^3E({R},\eta,0)}{\partial R \partial \eta^2}\right|_{0}\eta^{(1)})\\
\left.\frac{\partial^2 E({R},\eta,0)}{\partial \eta^2}\right|_{1}=
\left.\frac{\partial^2 E({R},\eta,0)}{\partial \eta^2}\right|_{0}
+{\cal E}_1(\left.\frac{\partial^3E({R},\eta,0)}{\partial R \partial \eta^2}\right|_{0}R^{(1)}
+\left.\frac{\partial^3E({R},\eta,0)}{\partial \eta^3}\right|_{0}\eta^{(1)})
\end{eqnarray}
Inserting all these results into Eq.~\ref{eq:tunability} completes the calculation of the linear dependence of $\chi_s$ on electric field.

\section{DISCUSSION}
\label{sec:discussion}


The expressions presented in the previous section are for the collinear tunability
of a material with a nonzero spontaneous polarization, where symmetry allows a linear
dependence of $\chi_s$ on $\cal E$. 
To obtain tunability of paraelectrics and noncollinear tunability of ferroelectrics,
where the linear term is zero by symmetry,
requires going one order higher in the electric field.
To this end, the necessary combinations of fourth-order energy derivatives can
be obtained from DFPT calculations of third derivatives by an extension of
the finite difference approach described above to obtain
the necessary combinations of third derivatives from calculated second
derivatives. As in the latter calculation, the highest derivatives of the
energy with respect to field cannot be obtained by finite difference calculations
in the structure, and have to be estimated. Such approximations are 
acceptable in light of the fact that the lattice contribution to the dielectric
constant dominates.

Extension to higher order is also of interest for collinear tunability of ferroelectrics, 
as measurements in films allow the determination of the electric field dependence of $\chi_s$
up to very high fields, showing strongly nonlinear behavior \cite{KN} that cannot be
investigated using the relatively low-order expansions of the present work.
The development of methods for density functional calculations in finite fields
\cite{Souza} will make it possible to investigate this regime.
However, for the low-field tunability, the method described here should be 
practical for even fairly complex-structured materials.

\section{ACKNOWLEDGMENTS}
This work was supported by NSF/NIRT DMR-0103354 and NSF/MRSEC DMR-0080008. K.M.R. thanks C. Fennie, J. B. Neaton, N. Sai, D. Schlom and D. Vanderbilt for valuable discussions.

%

\begin{thebibliography}{0}
\bibitem{nostrain}
It should be noted that the computation of first order wavefunctions and
associated derivatives for the homogeneous strain perturbation is not 
implemented in the present version of ABINIT.
\end{thebibliography}
 \end{document}